\documentstyle[12pt]{article}
\title{\bf Effect of the Flavor Changing Neutral Current on
Rare $B$ decays}
\author{L.T. Handoko\thanks{On leave of absence from Applied
	Physics Research and Development Center -~ Indonesian Institute of
	Sciences (PFT-LIPI), Serpong-INDONESIA.	E-mail address :
	handoko@theo.phys.sci.hiroshima-u.ac.jp} and T. Morozumi\thanks{
	This talk is presented by T. Morozumi}}
\date{\small \it Department of Physics, Hiroshima University \\
	1-3-1 Kagamiyama, Higashi Hiroshima - 724, Japan}

\begin{document}
\setlength{\baselineskip}{14pt}
\renewcommand{\thesubsection}{\arabic{subsection}}
\maketitle

\begin{abstract}
	We study the effect of the FCNC on rare $B$ decays in the
        beyond standard model with vector like singlet quarks.
        It has been shown that $b \rightarrow s \gamma $ does not
        receive sizable contribution compared with that of the standard
        model while $b \rightarrow d \gamma $, $b \rightarrow s(d) l^+ l^-$,
	can be changed from the predictions of the standard model.
\end{abstract}

\subsection{\bf Introduction}

\hspace*{1em} It has been known that the Flavor Changing process in
down quark sector is sensitive to the mass difference among
up-type quark in the standard model (SM). GIM mechanism tells us that
the FCNC in one-loop level of SM vanishes when up type
quarks are degenerate. In the real world , there is  large mass gap
among up quark sector ($M_u \ll M_c \ll M_t$).
Then the Flavor Changing Process in down quark sector like $b \rightarrow
s \gamma$ and $b \rightarrow s l^+ l^-$ are  enhanced. If we extend the
fermion sector beyond the standard model, there are several different
possibilities. If we just add the chiral fermions
as fourth generation in sequential way, GIM mechanism still
works and the new contribution
comes from up-type quark ($t^{\prime}$) in the fourth generation. Therefore
we can get constraints
for the mass of $t^\prime$ and its Kobayashi Maskawa mixing to
light down-type quarks ($V_{CKM}^{t^\prime i}$, $i = d,s,b$)
However if we extend the fermion
sector in non-sequential way, the different aspect
arises (refs. \cite{branco}, \cite{nir}, \cite{morozumi},
\cite{gautam} and \cite{handoko}).
The tree level FCNC arises both in $Z$ and neutral Higgs sector.
Furthermore, the size of the FCNC depends on the structure of
the down quark mass matrix rather than up type quark masses.
Therefore we may study the structure of down type quark mass
matrix by studying the flavor changing process in down quark sector.

\subsection{\bf Non-sequential extention of fermion sector}

\hspace*{1em} We study the Standard Model (SM) with extended quark sector.
In addition to the three standard generations of quarks,
$N_d - 3$ down-type and $N_u - 3$ up-type vector like singlet quarks are
introduced (refs. \cite{branco}, \cite{nir} and \cite{morozumi}).
\begin{equation}
	{I_W} = {1\over 2}\: : \:
	\left(
	\begin{array}{c}
		u \\
		d
	\end{array} \right)_L
	\left(
	\begin{array}{c}
		c \\
		s
	\end{array} \right)_L
	\left(
	\begin{array}{c}
		t \\
		b
	\end{array} \right)_L ,
	I_W = 0 \: : \:
	\left(
	\begin{array}{ccccc}
		u_R & c_R & t_R & t^{\prime}_{L+R} & \cdots \\
          	d_R & s_R & b_R & b^{\prime}_{L+R} & \cdots
       \end{array}\right) \: .
\end{equation}
In this model, there is tree level FCNC in $Z$ and Higgs sector.
In order to illustrate this point and  explain the relation between the
size of FCNC and down-type quark mass matrix, let us introduce
a toy model. This is the so called "top-prime less" model
\begin{equation}
	{I_W} = {1\over2} \: : \:
 	\left(
	\begin{array}{c}
		{t}^0 \\
		{b}^0
	\end{array}
	\right)_L \: , \:
	{I_W} = 0 \: : \:
 	\begin{array}{c}
		{t_R}^0 \\
		{b_R}^0
	\end{array}
 	{b_{L+R}}^{0\prime} \: .
\end{equation}
The most general mass matrix for down quark sector $M_d$ is given by,
\begin{equation}
 	L_{mass} = -(\bar{{b_L}^0} \: \bar{{{b_L}^0}^{\prime}})
	\left[
	\begin{array}{cc}
       		m & 0 \\
      		J & m_4
	\end{array} \right]
 	\left(
	\begin{array}{c}
		{b_R}^0 \\
		{{b_R}^0}^{\prime}
	\end{array}\right) \: .
\end{equation}
where $J$ is complex and $m$ and $m_4$ are real numbers. $m$ comes from the
vacuum expectation value of Higgs doublet.
$J$ leads to the mixing between left handed singlet quarks and right handed
ordinary quarks. In the limit of $J = 0$, the vector like quark decouples
from the ordinary quark.
$ M_d {M_d}^\dagger$ can be diagonalized by the following unitary
transformation :
\begin{equation}
	\left( \begin{array}{c}
		{b_L}^0 \\
		{b_L}^{\prime 0}
	\end{array} \right) =
	\left(
	\begin{array}{cc}
		\cos \theta & - \sin \theta e^{-i \phi} \\
		\sin \theta e^{i \phi} & \cos \theta
	\end{array} \right)
	\left(\begin{array}{c}
		b_L \\
		b_L^{\prime}
	\end{array} \right) \: ,
\end{equation}
where $b^0$ and $b^{0\prime}$ indicate the weak basis.
When $m$ is much smaller than $M = \sqrt{|J|^2 + {m_4}^2}$,
the elements of the
unitary matrix are  given by the following formulae
approximately :
\begin{eqnarray}
	\cos \theta & \simeq & 1 \: , \\
	\sin \theta e^{-i \phi} & \simeq & {m J \over M^2} \: .
\end{eqnarray}
Correspondingly, the mass eigenvalues
for heavy quark $(M_H)$ and light quark $(M_L)$ are given
by :
\begin{eqnarray}
	M_L & = & m {m_4 \over M} \\
 	M_H & = & M \: .
\end{eqnarray}
There are two interesting limits :
\begin{enumerate}
	\item $J \ll m_4$
	\item $J \gg m_4$
\end{enumerate}
In the case of (1),  the mixing between ordinary quark and
vector like quark is suppressed. The physical masses are given by
the diagonal elements of the mass matrix:
\begin{equation}
	M_L = m \: , \:  M_H = m_4 \: .
\end{equation}
In the case of (2),
the diagonal elements no longer reflect the physical masses :
\begin{equation}
	M_L = m {m_4 \over |J|} \: , \: M_H = |J| \: .
\end{equation}
Note that the light quark mass vanishes in the limit
$m_4 = 0$. On the other hand,
the neutral current is written in terms of physical basis,
\begin{eqnarray}
	L_Z & = & -{g \over 2 \cos \theta_W} \bar{{b^0_L}}
		\gamma_{\mu} {b^0}_L Z^{\mu} \nonumber \\
    	& = & -{g \over 2 \cos \theta_W} \left[
		\cos^2\theta \bar{b_L} \gamma_{\mu} b_L
       		-\cos \theta \sin \theta e^{-i \phi}
 		\bar{b_L} \gamma_{\mu} b_L^{\prime} + h.c.
		\right]Z^{\mu} \: .
\end{eqnarray}
Thus the flavor diagonal coupling for neutral current ($Z^{bb}$)
and the FCNC coupling ($Z^{bb^\prime}$) are given by :
\begin{eqnarray}
	Z^{bb} & = &\cos^2\theta = 1 - \sin^2\theta \simeq
		1 - O\left( \left(m J \over M^2 \right)^2
		\right) \: , \\
	Z^{bb^\prime} & = & -\cos \theta \sin \theta e^{-i \phi}
		\simeq -{m J\over M^2} \: .
\end{eqnarray}
Depending on the two cases mentioned above, the enhancement
and suppression of the FCNC occur respectively.
\begin{enumerate}
\item $J \ll m_4$ (suppression)
	\begin{eqnarray}
		M_L & = & m \: , \\
		M_H & = & m_4 \: , \\
		\left| Z^{bb^{\prime}}\right| & = & {m |J| \over {m_4}^2}
		\ll {M_L \over M_H} \: .
	\end{eqnarray}
\item $J \gg m_4$ (enhancement)
	\begin{eqnarray}
		M_L & = & m {m_4 \over |J|} \: , \\
		M_H & = & \left| J \right| \: , \\
		\left| Z^{bb^{\prime}} \right| & = & {m \over |J|}
			\gg {M_L \over M_H} \: .
	\end{eqnarray}
\end{enumerate}
For instance, if the physical mass of the vector like quark mass
is $M_H = 500$(GeV), the following mass matrices realize
the same mass eigenvalues for heavy and light quark masses
($M_H = 500$(GeV), $M_L = M_b = 5$(GeV)) while giving rise
to the different size of FCNC.
\begin{enumerate}
	\item \( M_d = \left[
		\begin{array}{cc}
			5 & 0 \\
			50 & 500
		\end{array} \right] (GeV) \: , \:
		Z^{bb^\prime} = 10^{-3} \: , \)
	\item \( M_d = \left[
		\begin{array}{cc}
			50 & 0 \\
			500 & 50
		\end{array} \right] (GeV) \: , \:
		Z^{bb^\prime} = 10^{-1} \: . \)
\end{enumerate}
This exercise tells us that the FCNC between singlet quark
and ordinary quark is enhanced when the off-diagonal element
in the mass matrix is larger than the diagonal element.

This kind of analysis can be extended to the general case with arbitrary
numbers of isosinglet up-type quarks and down-type quarks.
Here we just record the full lagrangian for the model (ref.\cite{handoko}).
\begin{eqnarray}
	{\cal L} & = & {\cal L}_{W^{\pm}} + {\cal L}_{\chi^{\pm}} +
		{\cal L}_A + {\cal L}_Z + {\cal L}_H + {\cal L}_{\chi^0} ,
	\label{eqn:lagrangian}
\end{eqnarray}
where,
\begin{eqnarray}
	{\cal L}_{W^{\pm}} & = & \frac{g}{\sqrt{2}}
		V_{CKM}^{\alpha\beta} \bar{u}^{\alpha} \gamma^{\mu} \, L \,
		d^{\beta} \, W_{\mu}^+ + h.c. , \\
	{\cal L}_{\chi^{\pm}} & = & \frac{g}{\sqrt{2} M_W}
		V_{CKM}^{\alpha\beta}
		\bar{u}^{\alpha} \left( m_{u^{\alpha}} L -
		m_{d^{\beta}} R \right) d^{\beta} \chi^+ + h.c. , \\
	{\cal L}_A & = & \frac{e}{3} \left(
		2 \bar{u}^{\alpha} \gamma^{\mu} u^{\alpha}
		- \bar{d}^{\alpha} \gamma^{\mu} d^{\alpha}
		\right) \, A_{\mu} , \\
	{\cal L}_Z & = & \frac{g}{2\cos\theta_W}
		\left\{ \bar{u}^{\alpha} \gamma^{\mu} \left[\left(
		{Z_u}^{\alpha\beta} - \frac{4}{3} \sin^2\theta_W
		\delta^{\alpha\beta}
		\right) L - \frac{4}{3} \sin^2\theta_W \delta^{\alpha\beta} R
		\right] u^{\beta} \right. \nonumber \\
	& &	+ \left. \bar{d}^{\alpha} \gamma^{\mu}
		\left[\left( \frac{2}{3} \sin^2\theta_W \delta^{\alpha\beta} -
		{Z_d}^{\alpha\beta}\right) L + 							\frac{2}{3}\sin^2\theta_W
\delta^{\alpha\beta}
		R\right] d^{\beta}\right\} \, Z_{\mu} , \\
	{\cal L}_H & = & \frac{-g}{2 M_W} \left[
		{Z_u}^{\alpha\beta} \bar{u}^{\alpha}
		\left(m_{u^{\alpha}} L + m_{u^{\beta}} R\right) u^{\beta}
		\right. \nonumber \\
	& &	\left. \: \: \: \: \: \: \: \:
		+ {Z_d}^{\alpha\beta} \bar{d}^{\alpha}
		\left(m_{d^{\alpha}} L + m_{d^{\beta}} R\right) d^{\beta}
		\right] \, H ,  \\
	{\cal L}_{\chi^0} & = & \frac{-ig}{2 M_W} \left[
		{Z_u}^{\alpha\beta} \bar{u}^{\alpha}
		\left(m_{u^{\alpha}} L - m_{u^{\beta}} R\right) u^{\beta}
		\right. \nonumber \\
	& &	\left. \: \: \: \: \: \: \: \:
		- {Z_d}^{\alpha\beta} \bar{d}^{\alpha}
		\left(m_{d^{\alpha}} L - m_{d^{\beta}} R\right) d^{\beta}
		\right] \, \chi^0 .
\end{eqnarray}
For $N_u = 3$ and $N_d = 4$, the diagonal elements and off-diagonal elements
of FCNC are given by the following equations :
\begin{eqnarray}
	Z^{bb} & = & 1 - \left| {V_L}^{4b} \right|^2 \: , \\
	Z^{bs} & = & -{{V_L}^{4b}}^\ast V_L^{4s} \: , \\
	Z^{bb^\prime} & = & -{{V_L}^{4b}}^\ast V_L^{44} \: ,
\end{eqnarray}
where $V_L$ is a unitary matrix which diagonalizes
the down type quark mass matrix
\begin{equation}
	 {d_L}^{0a} = {V_L}^{ab} {d_L}^b \: .
\end{equation}
In the basis in which up-type quark is diagonalized, $3$ by  $4$ part of
$V_L$ is just CKM matrix,
\begin{equation}
	{V_L}^{i a} = V_{CKM}^{ia} \: \:
		(i = 1,2,3 \: , \:  a = 1,2,3,4) \: .
\end{equation}
This leads to the relation between the FCNC and the deviation of unitarity
of CKM matrix
\begin{equation}
	Z^{bs} = -{{V_L}^{4b}}^\ast V_L^{4s}
		= \sum_{i=1}^{3} {V_{CKM}^{ib}}^\ast V_{CKM}^{is} \: .
\end{equation}
Therefore the unitarity of the CKM matrix no longer holds.
This non-unitarity ``quadrangle'' relation is used to
constrain the FCNC coupling, e.g. $Z^{bs}$, in later section.
However,
the deviation is suppressed when  the diagonal element
of singlet quark mass is infinite and keeping off-diagonal
element finite.

\subsection{Rare decays and Effect of FCNC}

\subsubsection{\it New Physics v.s. Standard Model
in $b \rightarrow s Z \rightarrow s l^+ l^-$}

\hspace*{1em} In the present  beyond standard model, there
is FCNC which leads to
tree level coupling $Z^{bs}$, $Z^{bd}$ and $Z^{sd}$.
In the standard model, the same vertex comes from top quark one loop
diagrams. Therefore the condition that the $Z$ FCNC dominates over the
standard model contribution in $Z$ sector is :
\begin{equation}
	\left| {Z^{bs} \over {V_{CKM}^{tb}}^\ast V_{CKM}^{ts}}
	\right| > O(\alpha) \simeq 0.012 \: .
\end{equation}
Then,  even tiny coupling for $Z$ FCNC, it can easily
dominates over the standard model contribution.
Experimental constraints and quadrangle constraints for the
FCNC in $bs$, $bd$ and $sd$ sectors are given in Table (\ref{tab:bound}).
{}From Table (\ref{tab:bound}) (refs. \cite{nir},
\cite{data} and \cite{handoko}),
$\left|{Z^{bs}}/{{V_{CKM}^{tb}}^\ast
V_{CKM}^{ts}} \right| \le 0.05 $ and $ \left| {Z^{bd}}/
{{V_{CKM}^{tb}}^\ast V_{CKM}^{td}} \right| \le 0.79$.
Therefore, there are allowed regions for $Z^{bs}$ and
$Z^{bd}$
couplings where the tree level $Z$ FCNC can dominate over the
1-loop top quark diagrams in the standard model.
We can expect the drastic change of the differential
decay rates
in the present beyond standard model in $b \rightarrow s l^+
l^-$ and  $b \rightarrow d l^+ l^-$ process.
\begin{table}[t]
	\begin{center}
	\begin{tabular}{|l|l|l|} \hline
	Decay process & Experiment constraint & Quadrangle constraint \\
	\hline \hline
	$B \rightarrow X_s \mu^+ \mu^-$
		& $\left| \frac{Z^{bs}}{{V_{CKM}^{cb}}^\ast
			V_{CKM}^{cs}}\right|
		\le 0.047$
		& $\left| \frac{{V_{CKM}^{tb}}^\ast V_{CKM}^{ts}}
		{{V_{CKM}^{cb}}^\ast V_{CKM}^{cs}}\right| \ge 0.94$
	\\ \hline
	$B \rightarrow X_d \mu^+ \mu^-$
		& $\left| \frac{Z^{bd}}{{V_{CKM}^{cb}}^\ast
			V_{CKM}^{cd}}\right|
		\le 0.23$
		& $\left| \frac{{V_{CKM}^{tb}}^\ast V_{CKM}^{td}}
		{{V_{CKM}^{cb}}^\ast V_{CKM}^{cd}}\right| \ge 0.29$
	\\ \hline
	$K^+ \rightarrow \pi^+ \nu \bar{\nu}$
		& $\left| \frac{Z^{sd}}{{V_{CKM}^{cd}}^\ast
			V_{CKM}^{cs}}\right|
		\le 2.9 \times 10^{-4}$
		& $\left| \frac{{V_{CKM}^{td}}^\ast V_{CKM}^{ts}}
		{{V_{CKM}^{cd}}^\ast V_{CKM}^{cs}}\right| \ge 0$ \\
	\hline
	\end{tabular}
	\caption{Upper bound for $\left| Z^{\alpha\beta} \right| $ from
	experiments, with assuming $Z$ exchange tree diagrams are dominant.}
	\label{tab:bound}
	\end{center}
\end{table}

\subsubsection{\it New Physics v.s. Standard Model in
$b \rightarrow s \gamma$ and $b \rightarrow s g$ process}

\hspace*{1em} There is no flavor changing neutral currents
for $b \rightarrow s \gamma $ and $ b\rightarrow s g $ ($g$ :
gluon) processes in the beyond
standard model and standard model. Therefore, in order that
the new physics contribution dominates over the standard model contribution,
the FCNC coupling constant must be the same order as that of
the CKM matrix elements.
The condition for the new physics contribution dominates over the
standard model contribution is :
\begin{equation}
	\left| {Z^{bs} \over {V_{CKM}^{tb}}^\ast V_{CKM}^{ts}}
	\right| > O(1) \: .
\end{equation}
{}From Table (\ref{tab:bound}) (refs. \cite{nir},
\cite{handoko}) we can see that there will be no
significant contribution to $b \rightarrow s \gamma (g)$
process because $\left| {Z^{bs}}/{{V_{CKM}^{tb}}^\ast V_{CKM}^{ts}}
\right| < 0.05$.
For $b \rightarrow d \gamma (g)$, new physics contribution
can be significant because $\left| {Z^{bd}}/{{V_{CKM}^{tb}}^\ast V_{CKM}^{td}}
\right| < 0.79$.

Let us summarize the computation of $b \rightarrow d \gamma$
briefly. The amplitude is proportional to the magnetic
moment interaction (refs. \cite{inami}, \cite{gautam} and
\cite{handoko}),
\begin{equation}
	T = \frac{G_F e}{4 \sqrt{2} \pi^2} \; \bar{d_L} \sigma_{\mu \nu}
                b_R \epsilon^{\mu} q^{\nu} F \: ,
\end{equation}
where,
\begin{eqnarray}
	F & = & Q_u {V_{CKM}^{tb}}^\ast V_{CKM}^{td} F^{cc}
		\left( {m_t \over M_W} \right) \nonumber \\
 	& + & Q_d Z^{db} \left\{ {2 \over 3} \sin^2\theta_W
		{F_1}^{NC}(0) + 2 {F_3}^{NC}(0) \right\} \\
 	& + & Q_d Z^{db^\prime} Z^{b^\prime b} \left\{
		{F_3}^{NC}\left({m_{b^\prime} \over M_Z}\right)
		+ {F_2}^{NC}\left({m_{b^\prime} \over m_H},
		{m_{b^\prime} \over M_Z}\right) \right\} \nonumber
\end{eqnarray}
where $Q_u = 2/3$ and $Q_d=-1/3$, and
$m_H$ is the neutral Higgs mass.
The first term comes from the top quark loop.
The second and the third terms come from down type quark
loop. The second term represents the  light down type  quarks
$(d,b)$ loop and the third term represents the heavy down type quark
$(b^\prime)$. The details of the functions and their behavior
are given in ref.
\cite{handoko}.

\subsection{\bf Summary}

\hspace*{1em} We study the rare $B$ decays in the
non-sequential extention of
fermion sector. The effect of vector like down type singlet quark
is studied.
It has been shown that the tree level FCNC can contribute to
$ b \rightarrow s(d) l^+ l^-$ processes, while $b \rightarrow s
\gamma (g)$ does not receive the significant contribution due to
the New Physics. In $b \rightarrow d \gamma (g)$, the effect of
FCNC ($Z^{bd}$, $Z^{db^\prime}$, $Z^{bb^\prime}$) may be seen.
\bigskip
\noindent
{\large \bf Acknowledgments}

T. M.  would like to thank I. Bigi,
 J. Kodaira and T. Kouno for
comments. This work is supported by the
Grant-in-Aid for Scientific Research ($\sharp 06740220$) from the Ministry of
Education, Science and Culture, Japan. The work of L.T.H. was supported by
a grant from the Overseas Fellowship Program (OFP-BPPT), Indonesia.

\bigskip
\noindent
{\large \bf Q \& A}

\begin{description}
	\item[Q :I. Bigi]  I think that it is too optimistic to
	say that the $15$ percentage deviation from the SM can be
	seen in
	$b \rightarrow d \gamma$ process because of
	 $b \rightarrow u {\bar u} d \gamma$ process .
	\item[A :T. M. ]  I did not consider this point.
\end{description}

\end{document}